\documentclass[aps,prl,twocolumn,showpacs,superscriptaddress]{revtex4-1}  
\usepackage{graphicx}  
\usepackage{dcolumn}   
\usepackage{bm}        
\usepackage{amssymb}   

\begin{document}


\title{\textbf{Three-dimensional} foam flow resolved by fast X-ray tomographic microscopy}
\author{C. Raufaste} \affiliation{Universit\'e Nice Sophia Antipolis, CNRS, LPMC, UMR 7336, Parc Valrose, 06100 Nice, France}
\author{B. Dollet} \affiliation{Institut de Physique de Rennes, UMR CNRS 6251,
  Universit\'e de Rennes 1, Campus de Beaulieu, 35042 Rennes Cedex,
  France}
\author{K. Mader}
\affiliation{Institute for Biomedical Engineering, University and ETH Zurich, Gloriastrasse 35, Zurich, Switzerland}
 \affiliation{Swiss Light Source, Paul Scherrer Institute, Villigen, Switzerland}
\author{S. Santucci} \affiliation{Laboratoire de Physique, ENS Lyon, UMR CNRS 5672, 46 all\'ee d'Italie, 69007 Lyon, France}
\author{R. Mokso} \affiliation{Swiss Light Source, Paul Scherrer Institut, Villigen, Switzerland}


\begin{abstract}
Thanks to ultra fast and high resolution X-ray tomography, we managed to capture the evolution of the local structure of  the bubble network of a 3D foam  flowing around a sphere. As for the 2D foam flow around a circular obstacle, we observed an axisymmetric velocity field with a recirculation zone, and indications of a negative wake downstream the obstacle. The bubble deformations, quantified by a shape tensor, are smaller than in 2D, due to a purely 3D feature: the azimuthal bubble shape variation.
Moreover, we were able to detect plastic rearrangements, characterized by the neighbor-swapping of four bubbles. Their spatial structure suggest that rearrangements are triggered when films faces get smaller than a characteristic area.
\end{abstract}

\maketitle

Foam rheology is an active research topic \cite{Weaire1999,Cantat2013,Cohen-Addad2013,Dollet2014}, motivated by applications in ore flotation, enhanced oil recovery, food or cosmetics \cite{Stevenson2012}. Because foams are opaque, imaging their flow in bulk at the bubble scale is challenging. To bypass this difficulty, 
2D flows of foams confined as a bubble monolayer, which structure is easy to visualize, have been studied. However, the friction induced by the confining plates may lead to specific effects \cite{Wang2006}, irrelevant for bulk rheology. 
In 
3D, diffusive-wave spectroscopy has been used to detect plastic 
rearrangements \cite{Durian1991,Cohen-Addad2001}. These events, 
called T1s, characterized in 2D by the neighbor swapping of four bubbles in contact, are of key importance for flow rheology, since their combination leads to the plastic flow of foams. 
Magnetic resonance imaging has also been used to measure the velocity field in 3D \cite{Ovarlez2010}. However, both these techniques resolve neither the bubble shape, nor the network of liquid channels (Plateau borders, PBs) within a foam. In contrast, X-ray tomography renders well its local  structure.
However, the long acquisition time of a tomogram, over a minute until very recently, constituted its main limitation, allowing to study only slow coarsening processes \cite{Lambert2007,Lambert2010}.

Here, we report the first quantitative study of a 3D foam flow around an obstacle. Such challenge was tackled thanks to a dedicated ultra fast and high resolution imaging set-up, recently developed at the TOMCAT beam line of the Swiss Light Source \cite{Mokso2010}.  
High resolution tomogram covering  a volume of $4.8\times 4.8\times 5.6$ mm$^3$ with  a voxel edge length of 5.3 $\mu$m could be acquired in around 0.5 s, allowing to follow the evolving structure of the bubbles and PB network. Our image analysis shows that the 3D foam flow around a sphere is qualitatively similar to the  2D flow around a circular obstacle: we reveal an axisymmetric velocity field, with a recirculation zone around the sphere  in the frame of the foam, and a negative wake downstream the obstacle. Bubble deformations are smaller (in the diametral plane along the mean direction of the flow $z$) than for a 2D flow, thanks to the extra degree of freedom allowing an azimuthal deformation: bubbles appear oblate before, and prolate after, the obstacle.
Finally, we were able to detect plastic rearrangements, characterized by the neighbor-swapping of four bubbles and the exchange of two four-sided faces. Our observations suggest that those events are  triggered when the bubble faces  get smaller than a characteristic size around $R_c^2$, given by a cutoff length of the PB $R_c \simeq \ 130~\mu$m  in the case of our foam.
\begin{figure}
\includegraphics[width=7.5cm]{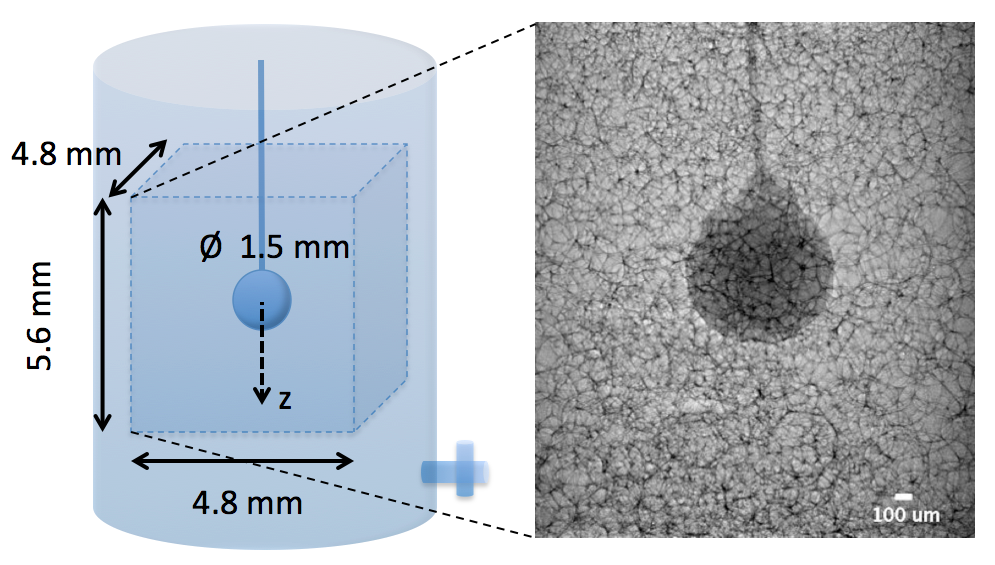}
\vspace{-0.5cm}
\caption{\label{fig:flow} A plastic bead of 1.5 mm diameter glued to a capillary is placed in the middle of the cylindric  chamber of 22 mm diameter and 50 mm height. The acquired tomograms cover the central region with a volume of $4.8\times 4.8\times 5.6$ mm$^3$. A typical X-ray projection image is shown on the right.}
\end{figure}

\emph{Experimental set-up --} 
We prepared a foaming solution following the protocol of \cite{Golemanov2008}: we mixed 6.6\% of sodium lauryl ether sulfate (SLES) and 3.4\% of cocamidopropyl betaine (CAPB) in mass in ultrapure water; we then dissolved 0.4\% in mass of myristic acid (MAc), by stirring and heating at 60$^\circ$C for one hour, and we diluted 20 times this solution. 
A few mL of solution was poured in the bottom of a cylindrical perspex chamber of diameter 22 mm and height 50 mm, compatible with 180 degree tomography. 
Bubbling air through a needle immersed in this solution, a foam was created until it reached the top of the chamber.
%
The bottom of the cell contains a tube connected to the open air and closed by a tap.
Controlling the opening of the tap, we could obtain a slow steady flow of the liquid foam. Its mean velocity determined a posteriori by image analysis is equal to $v_{\mathrm{flow}}=8~\mu$m/s. While flowing, the foam is deformed due to the presence of an obstacle, a smooth plastic bead of diameter 1.5 mm, attached to a capillary to fix its position in the middle of the chamber (Fig. \ref{fig:flow}).

The experiments were performed at the TOMCAT beamline of the Swiss Light Source. Filtered polychromatic X-rays with mean energy of 30 keV were incident on a custom made flow cell (Fig.~\ref{fig:flow}) attached to the tomography stage with three translational and a rotational degrees of freedom. The X-rays passing through the foam in the  chamber were converted to visible light by a 100~$\mu$m thick LuAG:Ce and detected by a 12 bit CMOS camera. Typically 550 radiographic projections acquired with 1 ms exposure time at equidistant angular positions of the sample were reconstructed into a 3D volume of $4.8\times 4.8\times 5.6$ mm$^3$ with isotropic voxel edge length of $ps=5.3~\mu$m. 
Such a 3D snapshot of the flowing foam 
is acquired in $t_{\mathrm{scan}}=0.55$~s, ensuring that motion artifacts are absent 
since $t_{\mathrm{scan}}<ps/v_{\mathrm{flow}}$. 
In order to follow the structural changes of the foam during its flow around the obstacle,
we recorded a tomogram every 35 seconds for approximately 20 minutes (resulting in around 36 tomograms).
The tomograms quality is enhanced using not only the X-rays attenuation by the sample, but also the phase shift of the partially coherent X-ray beam as it interacts with the PBs and senses the electron density variation in the sample \cite{Mokso2010}. 
This phase shift was retrieved using a single phase object approximation \cite{Paganin2002}.

\begin{figure}[t]
\includegraphics[width=7.5cm]{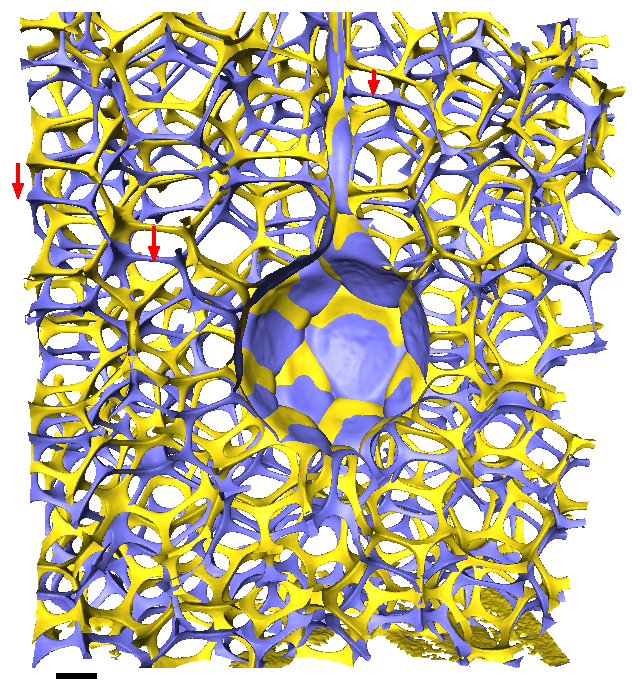}
\vspace{-0.3cm}
\caption{
\label{Fig:Skel}
3D volume representation of two instances in the foam flow. The PBs and vertices are colored in yellow and blue for time steps $t_0$ and $t_0+35$~s respectively. The scale bar is 300~$\mu$m. Red arrows indicate the flow direction.
}
\end{figure}
\emph{Image analysis --}
The tomograms are then segmented, separating the PBs and vertices from air.
Fig.~\ref{Fig:Skel} shows two successive time steps of the 3D snapshots of the reconstructed PBs network during the foam flow around the sphere.
We measured the liquid fraction from the segmented images, by computing the relative surface occupied by the PBs and vertices on individual horizontal slices. We measured an averaged liquid fraction of 4\% over a tomogram,
which did not evolve significantly during our experiments.  

%
Then, we reconstructed and identified individual bubbles of the flowing foam, following the procedure we recently developed and validated on static foam samples, imaged at the same acquisition rate and spatial resolution \cite{Mader2012}. We did not observe any evolution of the size distribution of the polydisperse foam studied here, with an average volume $V = 0.36 \pm 0.13$ mm$^3$, hence coarsening remains negligible. Typically, 160 bubbles are tracked between two successive 3D snapshots, leading to statistics over 5600 bubbles. 
To ensure a high tracking efficiency, we took into account the mean motion of the foam. 
Bubbles smaller than 0.01 mm$^3$ can't be discriminated from labeling artifacts \cite{Mader2012}, and thus,  are discarded.

\emph{Velocity field --}
From the bubble tracking, we could measure their velocity  around the obstacle.
Statistics are performed in the diametral ($rz$) plane of the cylindrical coordinates. We have checked that our results do not depend significantly on the angular coordinate $\theta$ (see also below), and we have averaged over this coordinate, as well as over time, thanks to the steadiness of the flow. The diametral plane is meshed into rectangular boxes ($0.25 \times 0.40$ mm$^2$). Consistently with the angular averaging procedure, we have checked that the number of bubbles analyzed per box is roughly proportional to the distance of the box to the symmetry axis (data not shown). Averages are weighted by the bubble volumes.
\begin{figure}
\includegraphics[width=7.5cm]{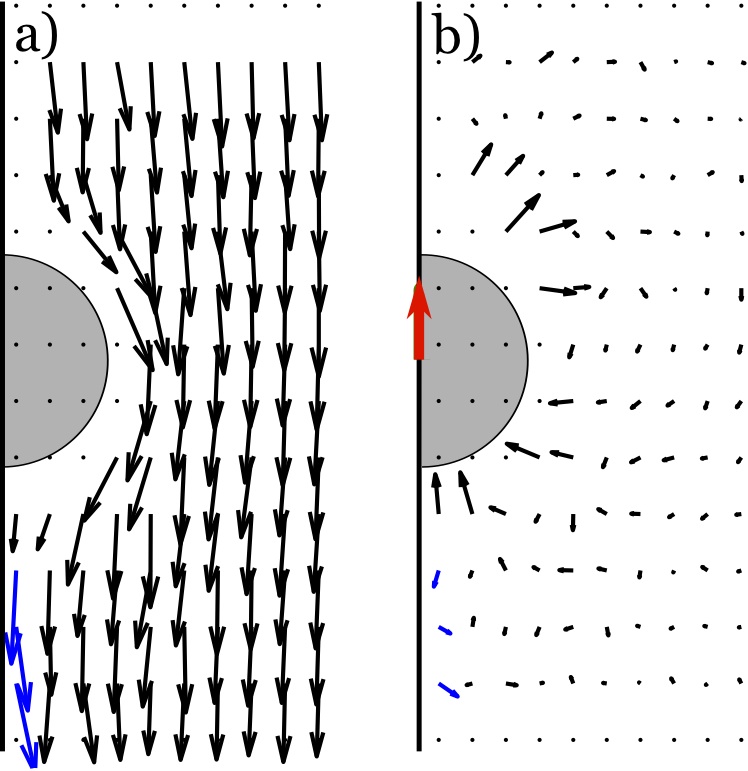}
\vspace{-0.3cm}
\caption{\label{fig:flowmap} Velocity fields in the ($rz$) plane, (a) in the lab frame. The normalized velocity field obtained by subtracting the mean flow velocity is shown in (b). The half-circle represents the obstacle (diameter 1.5~mm). The red arrow centered on the semi-obstacle gives the velocity scale of 8 $\mu$m/s.
Blue arrows show the negative wake effect.
}
\end{figure}
The velocity field is plotted in Fig.~\ref{fig:flowmap}.
In average over all patches, the $\theta$-component of the averaged velocity vector is  50 times smaller than its $rz$-component, hence the flow is axisymmetric. As expected, the velocity is uniform far from the obstacle, its amplitude decreases close to the leading and trailing points of the sphere, and increases along its sides. Accordingly, there is a clear recirculation zone surrounding the obstacle in the frame of the flowing foam (Fig.~\ref{fig:flowmap}b). It is worth noting that, compared to 2D foam flows around a circular obstacle \cite{Raufaste2007,Dollet2007,Marmottant2008,Cheddadi2011}, the range of influence of the obstacle on the flow field is  smaller. 

Interestingly, there is a zone downstream the obstacle and close to the symmetry axis where the streamwise velocity component is larger than the mean velocity or, equivalently, where the velocity opposes that of the obstacle in the frame of the flowing foam. This reminds the so-called \emph{negative wake}, revealed in viscoelastic fluids \cite{Hassager1979} and also evidenced in 2D foams \cite{Dollet2007}. However, a difficulty intrinsic to the 3D axisymmetric geometry is that the statistics is poor in these boxes close to the symmetry axis (about 10 bubbles per box over the full run), and should be improved in the future.

\begin{figure}
\includegraphics[width=6cm]{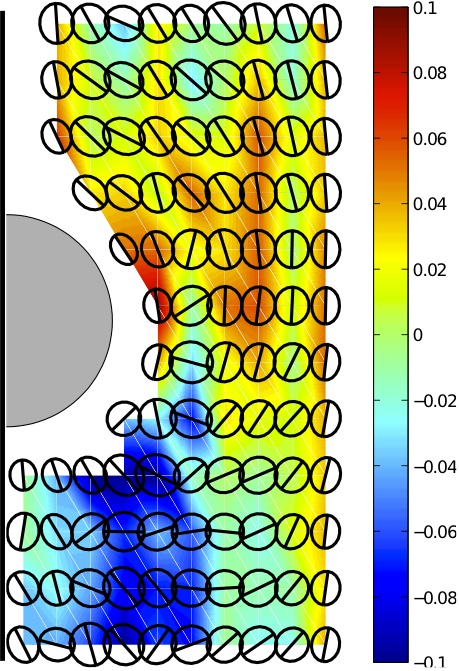}
\vspace{-0.3cm}
\caption{\label{fig:deformationmap} Projection of the bubble deformation field in the ($rz$) plane. Ellipses of bubbles are dilated by a factor of 3. The colormap gives the amplitude of the normalized deformation in the azimuthal direction.}
\end{figure}

\emph{Bubble deformation  --}
Given the set of coordinates $\{ \mathbf{r} \}$ of the voxels inside a bubble, we define its inertia tensor $\mathbf{I} = \langle (\mathbf{r} - \langle\mathbf{r}\rangle) \otimes (\mathbf{r} - \langle\mathbf{r}\rangle) \rangle$, and its shape tensor as $\mathbf{S} = \mathbf{I}^{1/2}$. This operation is valid because $\mathbf{I}$ (and hence $\mathbf{S}$) is a symmetric and definite tensor. The bubbles deformation is quantified by the eigenvectors/values of the shape tensor. In good approximation, two of them ($S_{rz}^+$ and $S_{rz}^-$) are found inside the ($rz$) plane, the other corresponds to the projection of the tensor along the azimuthal direction ($S_{\theta}$). An effective radius is defined by $R_{\mathrm{eff}} = (S_{rz}^+ S_{rz}^- S_\theta)^{1/3}$.
The bubbles deformation in the ($rz$) plane is represented by  ellipses of semi-axes $S_{rz}^+$ and $S_{rz}^-$. The direction of the largest one, $S_{rz}^+$, is emphasized by a line across the ellipse (Fig.  \ref{fig:deformationmap}). Deformation in the azimuthal direction is quantified by  $(S_\theta - R_{\mathrm{eff}})/R_{\mathrm{eff}}$ in colormap. The orientation of the ellipses in the ($rz$) plane exhibits a clear trend comparable to the 2D case \cite{Dollet2007,Graner2008}. They are elongated streamwise on the obstacle side and at the trailing edge. In between, the ellipses rotate $180^\circ$ to connect these two regions. We noticed that the deformation of the bubbles is much smaller than for a 2D foam with the same liquid fraction \cite{Dollet2007,Graner2008}. The deformation in the azimuthal direction exhibits dilation/compression up to 10\% only. The quantity $(S_\theta-R_{\mathrm{eff}})/R_{\mathrm{eff}}$ is positive upstream (oblate shape) to favor the passage around the obstacle (Fig.  \ref{fig:deformationmap}). The third dimension tends therefore to reduce the bubble deformation in the ($rz$) plane by increasing the deformation in the azimuthal direction. This effect is opposite downstream, right after the obstacle, where the bubbles are prolate.

\begin{figure}
\includegraphics[width=9cm]{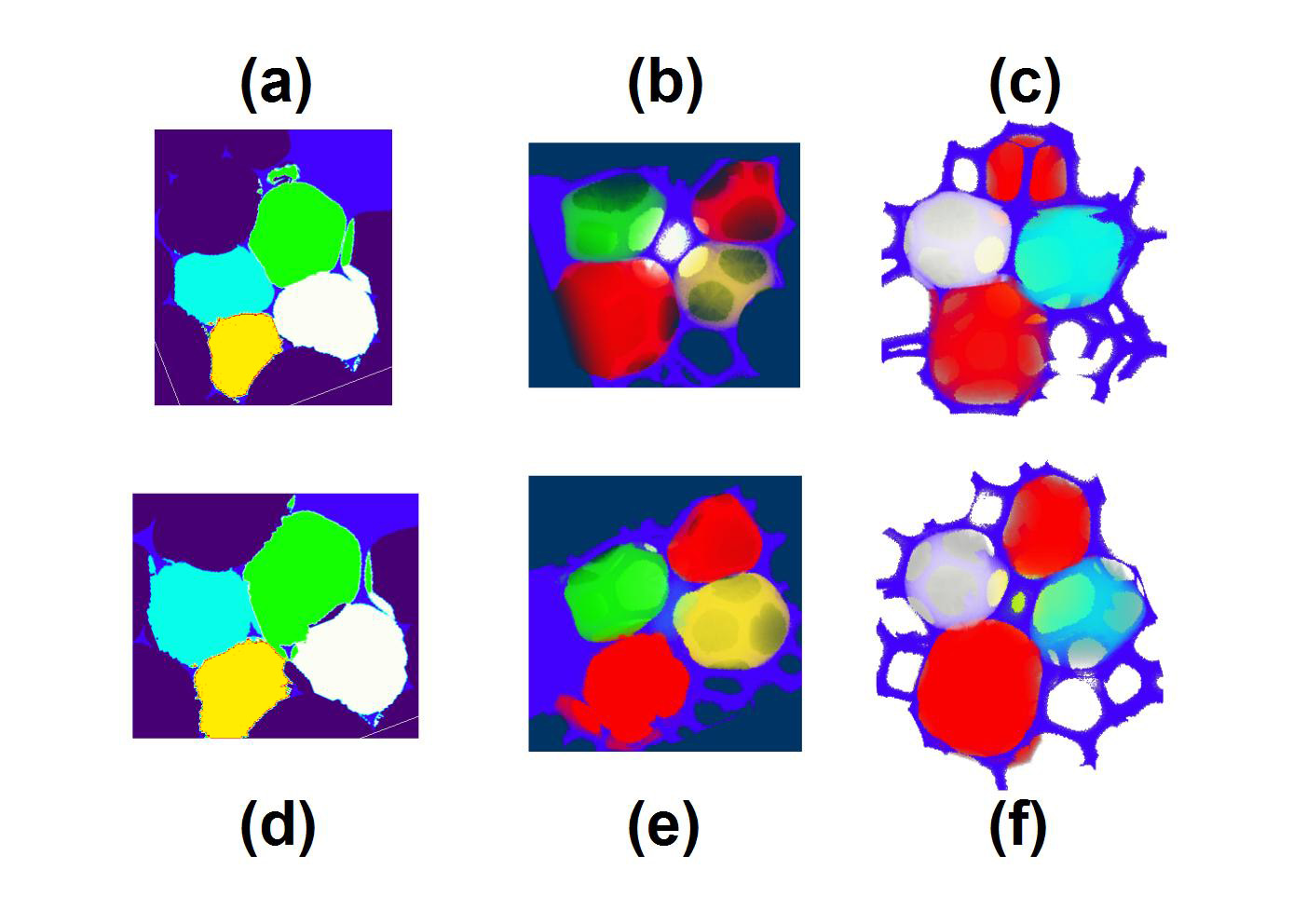}
\vspace{-0.6cm}
\caption{\label{fig:T1} Example of a T1 in 3D. The white and cyan bubbles lose contact, whereas the green and yellow bubbles come into contact. The red bubbles are the two bubbles that are in contact with these four bubbles over the rearrangement. Three different projections are shown: across the four bubbles that swap neighbors, (a) before and (d) after the T1; (b) in the plane of the face that is about to disappear, and (e) in this plane after the T1; (c) in the plane of the face that is about to appear and (f) in this plane after the T1.}
\end{figure}

\emph{Plastic rearrangements  --}
Automated tracking of bubble rearrangements was hindered by the high sensitivity of such procedure to small defects in the reconstruction of the bubble topology. Description of the contact between bubbles requires to rebuild precisely the faces between bubbles, which would require a finer analysis \cite{Davies2013}. Nevertheless, we managed to detect manually four individual events, corresponding to the rearrangements of neighboring bubbles. We provide below a detailed description of one typical example  (Fig.~\ref{fig:T1}); the features of the three other ones were found to be the same. Those  rearrangements consist of the swapping of four neighboring bubbles, with an exchange of four-sided faces, called T1s or quadrilateral-quadrilateral (QQ) transitions by Reinelt and Kraynik \cite{Reinelt1996,Reinelt2000}. We did not observed three-sided faces during a T1 as reported by \cite{Biance2009}. These are likely highly unstable, transient states which are too short-lived to be captured by tomography.
The QQ transitions observed involve two bubbles losing one face and two bubbles gaining one face. As can be seen on the projection plane across these four bubbles (Fig.~\ref{fig:T1}a and d), this is analogous to T1s in 2D, which always involve four bubbles, two losing one side and two gaining one side.
The distance between the two bubbles coming into contact decreases of $150 \ \mu$m, from 1.10~mm before the T1 to 0.95~mm after, while the distance between the two bubbles losing contact increases of $200 \ \mu$m, from 1.03~mm before the T1 to 1.23~mm after.
The distance between the two closest bubbles to the four ones directly involved in the T1 changes much less, from 1.48 to 1.52 mm.  This corroborates the vision of a T1 quite similar as in 2D, acting as a quadrupole in displacement, with most effect on the bubbles in the plane. On the other hand, the variation of shape anisotropy of the bubbles involved in the T1 did not show significant trends.

We went further on in the characterization of the spatial structure of those rearrangements. Bubble faces comprise a thin film surrounded by a thick network of PBs and vertices. All the faces that we observed are close to be isotropic, i.e. they do not show any significant elongation in a given direction. We have observed that the thin film part is usually very small for faces that are about to disappear, or that have just appeared, during a rearrangement. On the other hand, due to the finite radius of the PBs and of the finite size of the vertices, the ``skeleton" of these faces is not arbitrarily small. Quantitatively, we measured on the images a PB radius $R_{c} = 130$~$\mu$m. We also measured the area of the skeleton of the faces on 2D projections along the plane of the faces (we did not observe significantly non-planar faces). We always found skeleton areas larger than $3.4\times 10^4~\mu$m$^2$, which is of the order (a bit larger) of $R_{c}^2$. This suggests an interesting analogy with the cut-off edge length in 2D foams expected in theory \cite{Princen1983,Khan1989}, and measured in both simulations \cite{Cox2006} and experiments \cite{Raufaste2007}. In 2D foams and emulsions, when the distance between two approaching vertices reaches a certain length, a rearrangement occurs. This happens usually when the two PBs decorating the two neighboring vertices start to merge; hence, the order of magnitude of the cut-off length is $R_{c}$. For 3D foams, our observations suggest that there is a cut-off \emph{area} of the order of $R_{c}^2$ below which a face becomes unstable, triggering a rearrangement.

In summary, we have provided the first 
experimental measurement of a 3D time- and space-resolved foam flow measured directly from individual bubble tracking, with novel results on all the essential features of liquid foam mechanics: elasticity, plasticity and flow, through descriptions of shape field, T1 events, and velocity field. Such experimental results could be achieved  thanks to the recent advances of both high resolution and fast X-ray tomography and quantitative analysis tools.
Perspectives include further refinements of the reconstruction methods \cite{Mader2012,Davies2013}, to fully automatize the detection of rearrangements, to increase statistics and to study various geometries. Imaging the 3D flow at the bubble scale may shed new light on pending issues on shear localization \cite{Ovarlez2010} and nonlocal rheology \cite{Goyon2008}.

\begin{acknowledgments}
We thank Gordan Mikuljan from SLS who realized the experimental cells, Marco Stampanoni for supporting this project, the GDR 2983 Mousses et \'Emulsions (CNRS) for supporting travel expenses, Fran\c cois Graner, Gilberto L. Thomas and J\'er\^ome Lambert for discussions and the Paul Scherrer Institute for granting beam time to perform the experiments.
\end{acknowledgments}

\end{document}